\documentclass{svjour3}

\usepackage{graphics}
\usepackage{amssymb}
\usepackage{bm}
\usepackage{natbib}
\usepackage{fancyhdr}

\usepackage{latexsym}

\usepackage[abs]{overpic}

\usepackage{amsfonts}

\usepackage{amsmath}

\usepackage{mathrsfs}

\usepackage{graphicx}

\newcommand{\be}{\begin{eqnarray}}
\newcommand{\ee}{\end{eqnarray}}

\begin{document}

\title{A loophole of all `loophole-free' Bell-type theorems}

\author{Marek Czachor}

\institute{Katedra Fizyki Teoretycznej i Informatyki Kwantowej,
Politechnika Gda\'nska, 80-233 Gda\'nsk, Poland}
\maketitle

\begin{abstract}
Bell's theorem cannot be proved if complementary measurements have to be represented by random variables which cannot be added or multiplied. One such case occurs if their domains are not identical. The case more directly related to the Einstein-Rosen-Podolsky argument occurs if there exists an `element of reality' but nevertheless addition of complementary results is impossible because they are represented by elements from different arithmetics. A naive mixing of arithmetics leads to contradictions at a much more elementary level than the Clauser-Horne-Shimony-Holt inequality.
\end{abstract}


\section{Can we understand a theory that does not exist?}

When one speaks of a physical system that {\it violates\/} the Bell inequality \citep{Bell}, what one really has in mind is a system that {\it does not satisfy at least one assumption\/} needed for its proof. Some assumptions (locality of measurements, free will of observers,  perfect detectors) are physically quite obvious.  When it comes to postulating a joint probability measure for all random variables \citep{Vorobiev,Fine}, the situation is much less clear. Is it just synonymous to realism, another explicit postulate of Bell? Does it mean that counterfactual probabilities are identical to the measurable ones, even in cases where the alternative measurements cannot be simultaneously performed {\it in principle\/}, because of purely classical logical inconsistencies?

The latter is especially visible in the proof of the CHSH inequality \citep{CHSH}, which involves the following elementary step:
\be
a_0(x)b_0(y)+a_1(x)b_0(y)
=
\big(a_0(x)+a_1(x)\big)b_0(y).\label{1}
\ee
The value $b_0(y)$ of the random variable $b_0$, measured by Bob, does not depend on the choice of $a_0$ or $a_1$, measured by Alice. This is precisely the assumption of locality in the sense of Bell. In a nonlocal case we would have something like
\be
a_0(x,b_0)b_0(y,a_0)+a_1(x,b_0)b_0(y,a_1)\nonumber
\ee
so we could not further simplify the expression and proceed with the proof. The problem is easy to understand and is not a source of controversies.

However, the occurrence of $a_0(x)+a_1(x)$ in (\ref{1}) means that the sum of the two functions is well defined, which is not always the case in mathematics (think of $\ln x+\ln(-x)$). The subtlety is, of course, of a purely {\it local\/} nature. If a measurement of $a_0$ can influence in some way the one of $a_1$, the result of $a_0(x)+a_1(x)$ might in principle depend on the order in which the two measurements are performed, but addition is order independent. On the other hand, it can be shown on explicit examples (see \cite{MC92} which extends the original argument of \cite{Aerts1986}) that in cases where the measurements of $a_0$ and $a_1$ influence each other, the actual probabilities may differ from the counterfactual ones. One can prove a counterfactual inequality, but it may not apply to actually performed measurements. The attempts of finding a local counterexample to Bell's theorem \citep{Kupczynski} are, in my opinion, based on this loophole. 

The first version of the present paper was not meant for publication beyond arXiv.org. I just wanted to bring to a wider audience an old result of mine \citep{MC88}, whose importance seemed to grow, but which was virtually unknown. Traces of my old idea could be found in several recent papers \citep{Khrennikov,Christian2015,Luis},  but their authors were clearly unaware of my 1988 little contribution. However, once I started to think again of my old work, the subject started to live its own life. 

I believe the most important new element of the  paper is a reformulation of the CHSH inequality in the context of non-Diophantine arithmetic. It naturally leads to random variables whose addition or multiplication has to be done with great care. Since the goal of the Bell theorem is to eliminate all local hidden variable theories, the ones based on non-standard arithmetics should not be excluded. Non-Diophantine arithmetics and non-Newtonian calculus seem counterintuitive only at a first encounter. The are as natural as non-Euclidean geometry, non-Boolean logic, or non-Kolmogorovian probability.

We shall begin with a simple but rather formal example illustrating the main idea. Observers have free will, measurements are local, there are no undetected signals, and yet a Bell-type inequality cannot be proved for the same reason it cannot be proved in the nonlocal case. I will then proceed with a more subtle example based on non-Diophantine arithmetics. Here elements of reality in the sense of the EPR paradox are present, but the resulting complementary random variables cannot be added of multiplied, so that CHSH-type random variables cannot be automatically constructed.

\section{A formal example}

Alice and Bob are fans of two football teams, FC Aces and FC Bees, playing on Saturday nights in Acetown and Beetown. Whenever the Aces play in Acetown, Alice travels there and stays in a hotel. The hotels in Acetown  are named $A_\alpha$, where $0\leq \alpha\leq 1$. Alice is free to choose any of them, but typically stays at $A_0$ or $A_1$. Rooms in $A_\alpha$ are numbered by $x$, $\alpha< x <\alpha+1$. An analogous system works in Beetown. Unfortunately, the couple cannot travel together so if Alice supports her Aces in Acetown, then Bob is with the Bees in Beetown. Bob typically stays in $B_0$ or $B_1$, but is also completely free to make his choice. A peculiarity of the hotel system is that the same room can belong to several hotels (in the same town), a fact related to the structure of the local tax system, the antitrust law, and the unusual architecture typical of the region.

While leaving a hotel a visitor is asked to fill out a short questionnaire, reducing to a single question: Was it a nice visit? The answer is `$+$' or `$-$'. Experience shows that it is acceptable to stay in a room whose number is somewhere in the middle of the list:  $\alpha+1/4< x<\alpha+3/4$. Otherwise the noise made by fans who accompany the guest team becomes unbearable. Since decibels correspond to a logarithmic scale, the statistics of positive and negative answers in Acetown hotels is well described by the following random variable (Fig.~1 and Fig.~2):
\begin{figure}
\includegraphics[width=8 cm]{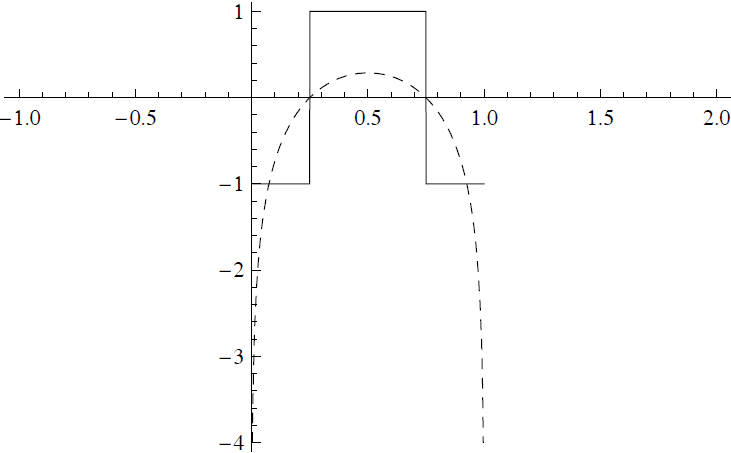}
\caption{Random variables $a_0(x)=\pm 1$ (full) and $\ln(16 x(1-x)/3)$ (dashed). $x$ is a room number in hotels $A_0$ or $B_0$. $x>1$ do not occur there and thus cannot be chosen by Alice or Bob.}
\end{figure}
\begin{figure}
\includegraphics[width=8 cm]{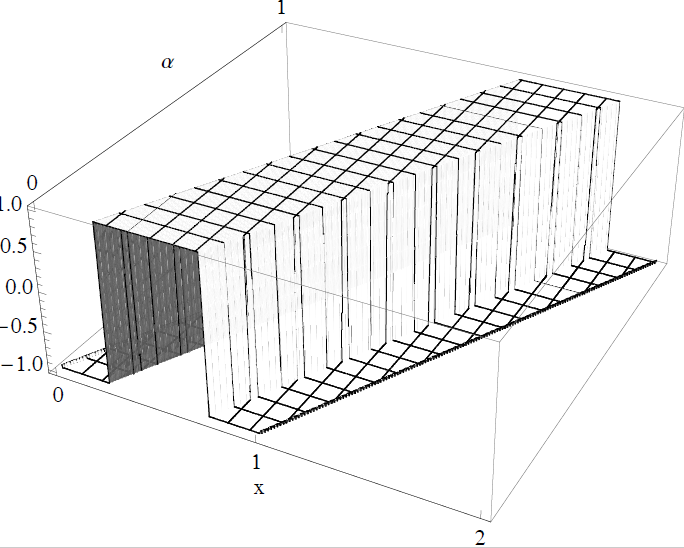}
\caption{Random variable $a_\alpha(x)$ as a function of $x$ and $\alpha$. For a given $\alpha$ the domain of the function is $(\alpha,\alpha+1)$.}
\end{figure}
\begin{figure}
\includegraphics[width=8 cm]{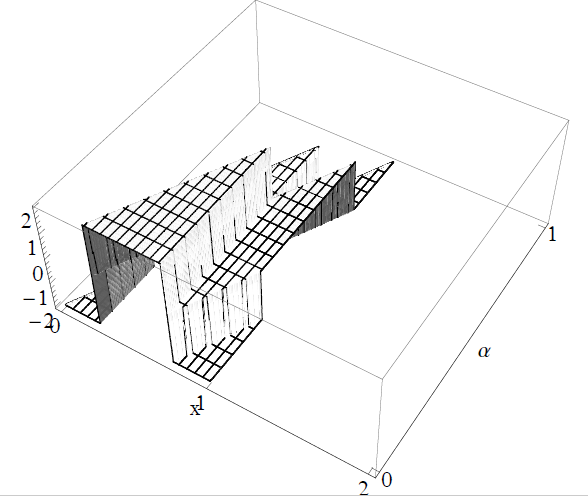}
\caption{Random variable $a_0(x)+a_\alpha(x)$ as a function of $x$ and $\alpha$. Its domain shrinks with growing $\alpha$ and becomes empty for $\alpha=1$.}
\end{figure}
\be
a_0(x)
&=&
\textrm{sign}\big(\ln(16 x(1-x)/3)\big),\\
a_\alpha(x)
&=&
a_0(x-\alpha).
\ee
The same model works in Beetown.

Each time Alice and Bob leave their hotels they fill out the forms and produce a pair of `results' $\pm$ (each run of the experiment produces a pair of results, hence no `detection loophole'). During their travels Alice and Bob do not communicate with each other (`locality'). Once one knows the room number, the result `$+$' or `$-$' is uniquely defined (`deterministic hidden variables'). Alice and Bob are free to choose the hotels (free will assumption). If Alice and Bob stayed in rooms with numbers $x$ in $A_\alpha$, and $y$ in $B_\beta$, then random variables $a_\alpha(x)$, $b_\beta(y)$, and $a_\alpha(x)b_\beta(y)$ are simultaneously defined.

So, can we prove Bell or CHSH inequalities? Let us try. Locality is satisfied, so apparently
\be
a_0(x)b_0(y)+a_1(x)b_0(y)
=
\big(a_0(x)+a_1(x)\big)b_0(y).
\ee
However, a look at Fig.~2 and Fig.~3 shows that $a_0(x)+ a_1(x)$ is undefined. The domain of the function is empty. It is not a zero function, but a function that does not exist at all. The putative proof got stuck already in its first line, in exactly the same way it gets stuck in the nonlocal case. The same will happen if instead of CHSH or Bell inequality one will try to prove the GHZ theorem \citep{GHZ}. Then, instead of the sum, one arrives at a product of the form $a_0(x)a_1(x)$, which does not exist either.

There is completely no problem with measuring experimentally an expectation value of $a_0(x)b_0(y)$, so there exists a probability distribution $\rho_{00}$ such that
\be
\langle a_0b_0\rangle
&=&
\int_0^1dx \int_0^1dy\, a_0(x)b_0(y)\rho_{00}(x,y).
\ee
The two functions, $a_0(x)b_0(y)$ and $\rho_{00}(x,y)$, are defined on the same domain $(0,1)\times (0,1)$.
Similarly,
\be
\langle a_1b_0\rangle
&=&
\int_1^2dx \int_0^1dy\, a_1(x)b_0(y)\rho_{10}(x,y),
\ee
with $a_1(x)b_0(y)$ and $\rho_{10}(x,y)$ defined on $(1,2)\times (0,1)$.

Note that the domains are disjoint, but there are no undetected signals. In 1988, pressed by a referee while resubmitting \citep{MC88}, I wrote that disjoint domains imply undetected signals (if $x\not\in (\alpha,\alpha+1)$ then $a_\alpha$ `produces no result'), although already at that time I was not quite convinced that the referee was right. Of course, one way of modeling non-ideal detectors is to use the trick with non-identical domains (this is what \cite{Pearle} does) but the inverse implication is not true. Another argument of the referee was that my random variables were three-valued: $\pm 1$ and 0, where zero occured when there `was no result' (again, this is what \cite{Pearle} assumes; the latter interpretation has been recently challenged by \cite{Christian19}). This type of argumenation does not apply here: Logarithm is not `zero' if its argument is negative. The same applies to $a_\alpha(x)$ for $x\not\in (\alpha,\alpha+1)$.

The only CHSH-type inequality one can find in our example is the trivial one,
\be
|
\langle a_0b_0\rangle
+
\langle a_1b_0\rangle
+
\langle a_0b_1\rangle
-
\langle a_1b_1\rangle
|\leq 4.\label{4}
\ee
Formally, having four different $\rho_{\alpha\beta}$, it is not difficult to give examples that saturate the right-hand side of (\ref{4}), while maintaining
\be
\langle a_0\rangle=\langle a_1\rangle=\langle b_0\rangle=\langle b_1\rangle=0.
\ee
An example of a theory that reproduces quantum probabilities, and which is implicitly based on the trick with disjoint domains of complementary observables can be found in Pitowsky's monograph \citep{Pitowsky}. When stripped of abstract details (decomposition of a unit interval into infinitely many {\it disjoint\/} non-measurable sets), it turns out that what is essential for the Pitowsky construction is the disjointness and not the non-measurability. So, instead of decomposing $[0,1]$ into non-measurable sets, one can replace $[0,1]$ by $[0,1]\times [0,1]$ and the effect is the same.

In the next section I will give further examples of random variables that cannot be carelessly added or multiplied.

\section{A subtler example: Non-Diophantine random variables}
\label{Sec 3}

One sometimes encounters physical quantities whose arithmetic is not the standard one. Velocities in special relativity are added by means of 
\be
\beta_1\oplus \beta_2=\tanh\big(\tanh^{-1}(\beta_1)+\tanh^{-1}(\beta_2)\big)
=f^{-1}\big(f(\beta_1)+f(\beta_2)\big),\label{beta}
\ee 
with $f(x)=\tanh^{-1}(x)$. Note that an analogous multiplication,
\be
\beta_1\odot \beta_2=f^{-1}\big(f(\beta_1)\cdot f(\beta_2)\big)=\tanh\big(\tanh^{-1}(\beta_1)\tanh^{-1}(\beta_2)\big)
\label{beta'},
\ee 
does not seem to occur in the literature. A parallel configuration of resistors is a resistor whose resistance is computed by means of the harmonic addition 
\be
R_1\oplus R_2 = 1/(1/R_1+1/R_2)=f^{-1}\big(f(R_1)+f(R_2)\big),\label{R}
\ee
with $f(x)=1/x$. Here multiplication is unchanged,
\be
R_1\odot R_2 =f^{-1}\big(f(R_1)\cdot f(R_2)\big)= 1/(1/R_1\cdot 1/R_2)=R_1R_2.\label{R'}
\ee
The rules (\ref{beta})--(\ref{R'}) can be regarded as particular examples of the so-called  projective non-Diophantine arithmetic, with `projection' $f$ and `coprojection' $f^{-1}$ \citep{Burgin77,Burgin,Burgin2,Burgin3},  constructed as follows.

Let $x, x'\in\mathbb{X}$ and assume there exists a bijection $f_\mathbb{X}:\mathbb{X}\to\mathbb{R}$ which defines all the four arithmetic operations in $\mathbb{X}$,
\be
x\oplus_\mathbb{X} x' &=& f_\mathbb{X}^{-1}\big(f_\mathbb{X}(x)+f_\mathbb{X}(x')\big),\\
x\ominus_\mathbb{X} x' &=& f_\mathbb{X}^{-1}\big(f_\mathbb{X}(x)-f_\mathbb{X}(x')\big),\\
x\odot_\mathbb{X} x' &=& f_\mathbb{X}^{-1}\big(f_\mathbb{X}(x)\cdot f_\mathbb{X}(x')\big),\\
x\oslash_\mathbb{X} x' &=& f_\mathbb{X}^{-1}\big(f_\mathbb{X}(x)/f_\mathbb{X}(x')\big).
\ee
Elements of $\mathbb{X}$ are ordered: $x\le_\mathbb{X} x'$ if and only if $f_\mathbb{X}(x)\le f_\mathbb{X}(x')$. 
The operations are commutative and associative, and $\odot_\mathbb{X}$ is distributive with respect to $\oplus_\mathbb{X}$. This is so because the one-to-one map $f_\mathbb{X}$ makes the arithmetic of $\mathbb{X}$ isomorphic to the standard Diophantine arithmetic of $\mathbb{R}$.

Let us stress that the only assumption we make about $\mathbb{X}$ is that its cardinality equals that of the continuum $\mathbb{R}$ (otherwise the bijection would not exist). $\mathbb{X}$ can be a highly nontrivial set, for example a fractal
\citep{MC2015,ACK2016a,ACK2016b,ACK2018,Czachor2019}, or a subset of $\mathbb{R}^n$.

The first two natural numbers, `zero' and `one' are the neutral elements of addition and multiplication. Denoting them by $0_\mathbb{X}$ and $1_\mathbb{X}$ we find
\be
0_\mathbb{X}=f_\mathbb{X}^{-1}(0),\quad 1_\mathbb{X}=f_\mathbb{X}^{-1}(1).
\ee
Indeed, $x\oplus_\mathbb{X} 0_\mathbb{X}=x$, $x\odot_\mathbb{X} 1_\mathbb{X}=x$, $x\ominus_\mathbb{X} x=0_\mathbb{X}$, and $x\oslash_\mathbb{X} x =1_\mathbb{X}$ (for $x\neq 0_\mathbb{X}$). A negative of $x$ is
$\ominus_\mathbb{X} x=0_\mathbb{X}\ominus_\mathbb{X} x=f_\mathbb{X}^{-1}\big(-f_\mathbb{X}(x)\big)$. An arbitrary natural number $n_\mathbb{X}$ is obtained by adding $1_\mathbb{X}$ an appropriate number of times,
\be
n_\mathbb{X}=\underbrace{1_\mathbb{X}\oplus_\mathbb{X}\dots\oplus_\mathbb{X} 1_\mathbb{X}}_{\textrm{$n$ times}}=f_\mathbb{X}^{-1}(n).
\ee
Then, in particular, $n_\mathbb{X}\oplus m_\mathbb{X}=(n+m)_\mathbb{X}$. An $n$-th power of $x$,
\be
x^{n_\mathbb{X}}=\underbrace{x\odot_\mathbb{X}\dots\odot_\mathbb{X} x}_{\textrm{$n$ times}}=f_\mathbb{X}^{-1}\big(f_\mathbb{X}(x)^n\big),
\ee
satisfies $x^{n_\mathbb{X}}\odot_\mathbb{X} x^{m_\mathbb{X}}=x^{n_\mathbb{X}\oplus_\mathbb{X} m_\mathbb{X}}$.  Finally, fractions are defined by $n_\mathbb{X}\oslash_\mathbb{X} m_\mathbb{X}$. For example
\be
(1_\mathbb{X}\oslash_\mathbb{X} 2_\mathbb{X})\oplus_\mathbb{X} (1_\mathbb{X}\oslash_\mathbb{X} 2_\mathbb{X})=1_\mathbb{X}=(1/2)_\mathbb{X}\oplus_\mathbb{X} (1/2)_\mathbb{X},
\ee
where $r_\mathbb{X}=f_\mathbb{X}^{-1}(r)$, now for any $r\in\mathbb{R}$. 

Probabilities are any numbers $p_j$ satisfying $0_\mathbb{X}\le_\mathbb{X}  p_j \le_\mathbb{X} 1_\mathbb{X}$ and
\be
\oplus_\mathbb{X}{}_{j=1}^n p_j=p_1\oplus_\mathbb{X}\dots\oplus_\mathbb{X} p_n=1_\mathbb{X}, 
\ee
which is equivalent to
\be
\sum_{j=1}^n f_\mathbb{X}(p_j)=1.
\ee
For example, $p_1=1_\mathbb{X}\oslash_\mathbb{X} 7_\mathbb{X}$, $p_2=2_\mathbb{X}\oslash_\mathbb{X} 7_\mathbb{X}$, $p_3=4_\mathbb{X}\oslash_\mathbb{X} 7_\mathbb{X}$ are probabilities.
Averages are defined analogously,
\be
\langle a\rangle
&=& \oplus_\mathbb{X}{}_{j=1}^n a_j\odot_\mathbb{X} p_j\\
&=&
f_\mathbb{X}^{-1}\Big(\sum_{j}f_\mathbb{X}(a_{j})f_\mathbb{X}(p_{j})\Big).\label{KN}
\ee
The form (\ref{KN}) is just the well known Kolmogorov-Nagumo average with respect to probabilities $P_{j}=f_\mathbb{X}(p_{j})$
\citep{K1930,N1930,CN,Naudts}

As a particular example of $f_\mathbb{X}$ consider $\mathbb{X}=\mathbb{R}$, $f_\mathbb{X}(x)=x^2\textrm{ sgn}(x)$, $f_\mathbb{X}^{-1}(x)=\sqrt{|x|}\textrm{ sgn}(x)$, $0_\mathbb{X}=0$, 
$(\pm 1)_\mathbb{X}=\pm 1$. Multiplication and division are unchanged. For $x\geq x'\geq 0$ we get
\be
x\oplus_\mathbb{X} x' &=& \sqrt{x^2+x'^2},\\
x\ominus_\mathbb{X} x' &=& \sqrt{x^2-x'^2},\\
x\odot_\mathbb{X} x' &=& x\cdot x',\\
x\oslash_\mathbb{X} x' &=& x/x'.
\ee
The average reads
\be
\langle a\rangle= \oplus_\mathbb{X}{}_{j=1}^n a_j\odot_\mathbb{X} p_j 
=
\pm 
\sqrt{\Big| \sum_j f(a_j)p_j^2\Big|},
\ee
where $\pm$ is the sign of $\sum_j f_\mathbb{X}(a_j)p_j^2$. It is clear that for this concrete form of arithmetic the non-Diophantine probabilities play essentially the role of real probability amplitudes.

Once one knows how to add, subtract, multiply, and divide, one can construct an entire calculus. Historically, the first example of such a `non-Newtonian calculus' was given by M.~Grossman and R.~Katz \citep{GK,G79,G83}. The idea was rediscovered by 
E.~Pap in his g-calculus \citep{Pap1993,Pap2008} and still later, but in its currently most general form, by myself 
\citep{MC2015,ACK2018,Czachor2019}. Certain old problems of fractal analysis (a Fourier transform on an arbitrary Cantor set, wave propagation along a Koch curve) have found simple solutions in the non-Newtonian framework \citep{ACK2016b,Czachor2019}.

Let us now consider a collection of different sets ${\mathbb{X}_k}$ equipped with different arithmetics, defined by different bijections $f_{\mathbb{X}_k}: {\mathbb{X}_k}\to \mathbb{R}$. To be concrete, let ${\mathbb{X}_1}=\mathbb{R}_+$ (positive reals), ${\mathbb{X}_2}=-\mathbb{R}_+$ (negative reals), $f_{\mathbb{X}_1}(x)=\ln x$, 
$f_{\mathbb{X}_1}^{-1}(r)=e^r$, $f_{\mathbb{X}_2}(x)=\ln (-x)$, $f_{\mathbb{X}_2}^{-1}(r)=-e^r$. 
There are reasons to believe that arithmetics of this type are employed by human and animal nervous systems \citep{C2020}. This is why decibels and star magnitudes correspond to logarithmic scales.

Assume that we have randomly selected a number $r\in\mathbb{R}$, say $r=1$. This is equivalent to randomly selecting 
$f_{\mathbb{X}_1}^{-1}(1)=1_{\mathbb{X}_1}=e\in \mathbb{X}_1$ and  
$f_{\mathbb{X}_2}^{-1}(1)=1_{\mathbb{X}_2}=-e\in \mathbb{X}_2$. Following the logic of the Bell theorem we could consider the sum (anyway, 
$1_{\mathbb{X}_1}$ and $1_{\mathbb{X}_2}$ are just real numbers),
\be
1_{\mathbb{X}_1}+1_{\mathbb{X}_2}=e+(-e)=0=f_{\mathbb{X}_1}^{-1}(-\infty)
=
\ominus_{\mathbb{X}_1}\infty_{\mathbb{X}_1}
, 
\ee
which, if correct, would  imply the inconsistent result
\be
e=-1_{\mathbb{X}_2}=1_{\mathbb{X}_1}\oplus_{\mathbb{X}_1}\infty_{\mathbb{X}_1}=\infty_{\mathbb{X}_1}
=f_{\mathbb{X}_1}^{-1}(\infty)=e^\infty=\infty.\label{1+1}
\ee
The inconsistency occurs because I have recklessly mingled three different arithmetics. 
The problem is that one is not allowed to naively add elements from $\mathbb{X}_1$ to those from $\mathbb{X}_2$  (see the next Section), so the whole calculation is meaningless. In spite of the fact that $1_{\mathbb{X}_1}+1_{\mathbb{X}_2}$ is zero, from the perspective of $\mathbb{X}_1$ or $\mathbb{X}_2$ the expression is ambiguous. Different arithmetics behave analogously to different maximal sets of jointly measurable quantities in quantum mechanics, an analogy worthy of  a separate study. There are arguments that problems with dark energy and dark matter may result from an analogous  mismatch of arithmetics \citep{C2017,C2020}. 

If the readers are not yet convinced, consider the case where $\mathbb{X}_1$ is the double cover of the Sierpi\'nski set discussed in \citep{ACK2018}, while $\mathbb{X}_2$ is the Cantor line from \citep{MC2015}. In the first case, $1_{\mathbb{X}_1}=(1,0)_+$ belongs to the positive side of an oriented $\mathbb{R}^2$. In the second case,
$1_{\mathbb{X}_2}=1\in\mathbb{R}$. What should be meant by their sum?

Very little is known about tensor products of arithmetics, an issue related to the difficult problem of tensor products of  fields (in the algebraic sense of this word). Even less is known about probabilistic correlation experiments involving measurements mathematically described by different arithmetics. Yet less is known about hidden-variable theories involving different arithmetics. The standard Bell theorem tells us virtually nothing about the subject.

So, let us try. But first a digression.

\section{Back to quantum mechanics (a digression)}

In quantum mechanics random variables are represented by self-adjoint operators. Values of the random variables are represented by eigenvalues of the operators. Commuting operators represent random variables that can be measured simultaneously. Tensor products of operators represent products of random variables associated with independent measurements. In the context of the CHSH inequality one deals with
\be
AB=\bm a\cdot\bm\sigma\otimes \bm b\cdot\bm\sigma,
\ee
where $\bm a$, $\bm b$ are unit vectors in $\mathbb{R}^3$. In general, the operator $\bm a\cdot\bm\sigma$ has eigenvalues $\pm |\bm a|$. This is why it is justified to treat $AB$ as a random variable whose values $\pm1$ are products of the eigenvalues $\pm1$ of $\bm a\cdot\bm\sigma$ and $\bm b\cdot\bm\sigma$. 

The problem with the Bell theorem is that  in the course of its proof one assumes that
\be
A+A' &=&  (\bm a\cdot\bm\sigma+\bm a'\cdot\bm\sigma)\otimes\mathbb{I}\label{A+A'},\\
A-A' &=&  (\bm a\cdot\bm\sigma-\bm a'\cdot\bm\sigma)\otimes\mathbb{I}\label{A-A'},
\ee
represent random variables whose values are 0,  $\pm 2$. However, since the eigenvalues of (\ref{A+A'}) and (\ref{A-A'}) are $\pm|\bm a+\bm a'|$ and $\pm|\bm a-\bm a'|$ the observables are two-valued and not three-valued, so 0  is not an eigenvalue unless $\bm a'=\pm\bm a$. A contradiction with quantum mechanics is obtained already at this stage. Nonlocality is not needed. One can also show that a quantum CHSH inequality holds  if and only if both $[A,A']$ and $[B,B']$ are nonzero. This follows from the fact \citep{Wolf,Khrennikov2019} that the Bell operator
\be
C=A\otimes B+A\otimes B'+A'\otimes B-A'\otimes B'
\ee
violates the inequality if and only if 
\be
C^2=4+[A,A']\otimes [B,B']> 4
\ee
and this is possible if both commutators are nonzero (effectively, if operators $B$, $B'$ with eigenvalues $\pm 1$ commute, then $B'=\pm B$, and two of the four terms in $C$ cancel out; then the argument on eigenvalues of $A\pm A'$ does not work since either $A$ or $A'$ is missing). The conditions $[A,A']\neq 0$ and 
$[B,B']\neq 0$ are, of course, of a local nature. The argument based on noncommutativity does not have an exact analogue in the arithmetic framework and is a peculiarity of the quantum formalism.

A similar difficulty will nevertheless occur.

\section{Arithmetics of Alice, Bob, and Eve}

Let us begin with the simplest problem of adding two averages. Assume Alice measures two random variables, $a$ and $a'$, with values in arithmetics $\mathbb{A}$ and $\mathbb{A}'$, respectively. After $n$ measurements of $a$ and $n'$ measurements of $a'$ the results are
$(a_1,\dots,a_n)\subset \mathbb{A}\times \dots \times \mathbb{A}$, $(a'_1,\dots,a'_{n'})\subset \mathbb{A}'\times \dots \times \mathbb{A}'$.  She computes averages,
\be
\langle a\rangle
&=& \oplus_\mathbb{A}{}_{j=1}^n a_j\oslash_\mathbb{A} n_\mathbb{A},\\
\langle a'\rangle
&=& \oplus_{\mathbb{A}'}{}_{j=1}^{n'} a'_j\oslash_{\mathbb{A}'} n'_{\mathbb{A}'}.
\ee
What should be meant by their sum, even if the experimental samples have equal length $n=n'$? 
Can we write $\langle a\rangle+\langle a'\rangle=\langle a+ a'\rangle$? In general, no.

The problem remains even if 
$\mathbb{A}\subset \mathbb{R}$ and $\mathbb{A}'\subset \mathbb{R}$. The random variables are real, but rules of their addition and division are different.
More strikingly, even if $a_j=r_{\mathbb{A}}=f^{-1}_{\mathbb{A}}(r)$, $a'_j=r'_{\mathbb{A}'}=f^{-1}_{\mathbb{A}'}(r)$, so that $r\in\mathbb{R}$ behaves analogously to an `element of reality' common to $a_j$ and $a'_j$, the sum of $a_j$ and $a'_j$ is as ambiguous as (\ref{1+1}). This seems to be the first example (outside of quantum mechanics) where one cannot add two random variables even though they may be regarded as possessing a common element of reality in the sense of the EPR argument.

Let us go further. Let Alice work with arithmetics $\mathbb{A}$ or $\mathbb{A}'$, and Bob with $\mathbb{B}$ or $\mathbb{B}'$.
They perform binary measurements with the results $1_{\mathbb{X}}$ or $\ominus_{\mathbb{X}}1_{\mathbb{X}}$, where 
$\mathbb{X}$ is the corresponding arithmetic. Although numbers can be consistently added or multiplied only within a single arithmetic, results from one arithmetic can be uniquely translated into another arithmetic by means of the  isomorphism with $\mathbb{R}$. For example, if
\be
x\oplus_\mathbb{A}^\mathbb{B} x' &=& f_\mathbb{A}^{-1}\big(f_\mathbb{A}(x)+f_\mathbb{B}(x')\big),\\
x\ominus_\mathbb{A}^\mathbb{B} x' &=& f_\mathbb{A}^{-1}\big(f_\mathbb{A}(x)-f_\mathbb{B}(x')\big),\\
x\odot_\mathbb{A}^\mathbb{B} x' &=& f_\mathbb{A}^{-1}\big(f_\mathbb{A}(x)\cdot f_\mathbb{B}(x')\big),\\
x\oslash_\mathbb{A}^\mathbb{B} x' &=& f_\mathbb{A}^{-1}\big(f_\mathbb{A}(x)/f_\mathbb{B}(x')\big).
\ee
then 
\be
2_\mathbb{A}\oplus_\mathbb{A}^\mathbb{B} 2_\mathbb{B} &=& f_\mathbb{A}^{-1}\big(2+2\big)=4_\mathbb{A},\\
2_\mathbb{A}\oplus_\mathbb{B}^\mathbb{A} 2_\mathbb{B} &=& f_\mathbb{B}^{-1}\big(2+2\big)=4_\mathbb{B}.
\ee
In the logarithmic example,  ${\mathbb{A}}=\mathbb{R}_+$, ${\mathbb{B}}=-\mathbb{R}_+$, $f_{\mathbb{A}}(x)=\ln x$, 
$f_{\mathbb{A}}^{-1}(r)=e^r$, $f_{\mathbb{B}}(x)=\ln (-x)$, $f_{\mathbb{B}}^{-1}(r)=-e^r$, we will of course conclude that `two plus two equals four', but the result looks as follows:
\be
2_\mathbb{A}\oplus_\mathbb{A}^\mathbb{B} 2_\mathbb{B} &=& e^4=4_\mathbb{A}
=
2_\mathbb{A}\odot_\mathbb{A}^\mathbb{B} 2_\mathbb{B},\\
2_\mathbb{A}\oplus_\mathbb{B}^\mathbb{A} 2_\mathbb{B} &=& -e^4=4_\mathbb{B}
=
2_\mathbb{A}\odot_\mathbb{B}^\mathbb{A} 2_\mathbb{B}.
\ee
Alice and Bob would agree that they have obtained identical results, namely `four', but Eve might disagree. In her opinion  their results were  exactly opposite: $e^4$ and $-e^4$.

The other two arithmetics needed for a CHSH random variable are, in principle, completely independent. For example, 
${\mathbb{B}'}=\mathbb{R}_+$, ${\mathbb{A}'}=-\mathbb{R}_+$, $f_{\mathbb{B}'}(x)=(\ln x)^3$, 
$f_{\mathbb{B}'}^{-1}(r)=e^{r^{1/3}}$, $f_{\mathbb{A'}}(x)=\big(\ln (-x)\big)^3$, $f_{\mathbb{A'}}^{-1}(r)=-e^{r^{1/3}}$.

Now, consider the following arithmetic operations
\be
a\oplus_\mathbb{X}^\mathbb{AB} b &=& f_\mathbb{X}^\mathbb{AB}{}^{-1}\big(f_\mathbb{A}(a)+ f_\mathbb{B}(b)\big),\\
a\ominus_\mathbb{X}^\mathbb{AB} b &=& f_\mathbb{X}^\mathbb{AB}{}^{-1}\big(f_\mathbb{A}(a)- f_\mathbb{B}(b)\big),\\
a\odot_\mathbb{X}^\mathbb{AB} b &=& f_\mathbb{X}^\mathbb{AB}{}^{-1}\big(f_\mathbb{A}(a)\cdot f_\mathbb{B}(b)\big),\\
a\oslash_\mathbb{X}^\mathbb{AB} b &=& f_\mathbb{X}^\mathbb{AB}{}^{-1}\big(f_\mathbb{A}(a)/ f_\mathbb{B}(b)\big),
\ee
where $a\in \mathbb{A}$, $b\in \mathbb{B}$, but $\mathbb{X}$ and the bijection $f_\mathbb{X}^\mathbb{AB}:\mathbb{X}\to\mathbb{R}$ are yet unspecified. 
Recall that for $r\in\mathbb{R}$ we denote $r_\mathbb{A}=f_\mathbb{A}^{-1}(r)$, etc. 
The product has the following properties:
\be
x_\mathbb{A}\odot_\mathbb{X}^\mathbb{AB} y_\mathbb{B} &=& f_\mathbb{X}^\mathbb{AB}{}^{-1}(xy)
=
(xy)_\mathbb{A}\odot_\mathbb{X}^\mathbb{AB} 1_\mathbb{B}
=
1_\mathbb{A}\odot_\mathbb{X}^\mathbb{AB} (xy)_\mathbb{B}\nonumber\\
&=&
(x_\mathbb{A}\odot_\mathbb{A}y_\mathbb{A})\odot_\mathbb{X}^\mathbb{AB} 1_\mathbb{B}
=
1_\mathbb{A}\odot_\mathbb{X}^\mathbb{AB} (x_\mathbb{B}\odot_\mathbb{B}y_\mathbb{B}),\\
x_\mathbb{A}\odot_\mathbb{X}^\mathbb{AB} (y_\mathbb{B}\oplus_\mathbb{B}z_\mathbb{B}) 
&=& f_\mathbb{X}^\mathbb{AB}{}^{-1}\big(x(y+z)\big)\nonumber\\
&=&
f_\mathbb{X}^\mathbb{AB}{}^{-1}\big(f_\mathbb{A}(x_\mathbb{A})f_\mathbb{B}(y_\mathbb{B})+f_\mathbb{A}(x_\mathbb{A})f_\mathbb{B}(z_\mathbb{B})\big)\nonumber\\
&=&
f_\mathbb{X}^\mathbb{AB}{}^{-1}\Big(
f_\mathbb{A}\circ f_\mathbb{A}^{-1}\big(f_\mathbb{A}(x_\mathbb{A})f_\mathbb{B}(y_\mathbb{B})\big)
+
f_\mathbb{B}\circ f_\mathbb{B}^{-1}\big(f_\mathbb{A}(x_\mathbb{A})f_\mathbb{B}(z_\mathbb{B})\big)\Big)\nonumber\\
&=&
f_\mathbb{X}^\mathbb{AB}{}^{-1}\Big(
f_\mathbb{A}\big(x_\mathbb{A}\odot_\mathbb{A}^\mathbb{B} y_\mathbb{B}\big)
+
f_\mathbb{B}\big(x_\mathbb{A}\odot_\mathbb{B}^\mathbb{A}z_\mathbb{B}\big)\Big)\nonumber\\
&=&
(x_\mathbb{A}\odot_\mathbb{A}^\mathbb{B} y_\mathbb{B})
\oplus_\mathbb{X}^\mathbb{AB}
(x_\mathbb{A}\odot_\mathbb{B}^\mathbb{A}z_\mathbb{B})
\ee
Since 
\be
x_\mathbb{A}\odot_\mathbb{X}^\mathbb{AB} y_\mathbb{B}=y_\mathbb{A}\odot_\mathbb{X}^\mathbb{AB} x_\mathbb{B},
\ee
we immediately also get 
\be
(x_\mathbb{A}\oplus_\mathbb{A}y_\mathbb{A})\odot_\mathbb{X}^\mathbb{AB} z_\mathbb{B} 
&=&
(x_\mathbb{A}\odot_\mathbb{A}^\mathbb{B} z_\mathbb{B})
\oplus_\mathbb{X}^\mathbb{AB}
(y_\mathbb{A}\odot_\mathbb{B}^\mathbb{A}z_\mathbb{B})
\ee
These are all the properties needed for a tensor product. 
Still, how should we add 
$a_\mathbb{A}\odot_\mathbb{X}^\mathbb{AB} b_\mathbb{B}\in\mathbb{X}$
and
$a_\mathbb{A}\odot_\mathbb{X}^\mathbb{AB'} b'_{\mathbb{B}'}\in\mathbb{X}$? The most natural addition is the one intrinsic to $\mathbb{X}$,
\be
(a_\mathbb{A}\odot_\mathbb{X}^\mathbb{AB} b_\mathbb{B})
\oplus_\mathbb{X}
(a_\mathbb{A}\odot_\mathbb{X}^\mathbb{AB'} b'_{\mathbb{B}'})
&=&
f_\mathbb{X}^{-1}
\Big(
f_\mathbb{X}(a_\mathbb{A}\odot_\mathbb{X}^\mathbb{AB} b_\mathbb{B})
+
f_\mathbb{X}(a_\mathbb{A}\odot_\mathbb{X}^\mathbb{AB'} b'_{\mathbb{B}'})
\Big)\\
&=&
f_\mathbb{X}^{-1}
\Big(
f_\mathbb{X}\circ f_\mathbb{X}^\mathbb{AB}{}^{-1}(ab)+f_\mathbb{X}\circ f_\mathbb{X}^\mathbb{AB'}{}^{-1}(ab')
\Big).
\ee
It cannot be further simplified, since $f_\mathbb{X}\circ f_\mathbb{X}^\mathbb{AB}{}^{-1}$ and $f_\mathbb{X}\circ f_\mathbb{X}^\mathbb{AB'}{}^{-1}$
are in general different functions, so we are back to our original argument. Bell's theorem cannot be proved if inconsistent sets of data have to be combined in a nontrivial way.

Let us now consider two extreme cases: (i) $\mathbb{X}=\mathbb{R}$, $f_\mathbb{X}(x)=x$, and (ii) a  nontrivial $\mathbb{X}$ but 
$f_\mathbb{X}^\mathbb{AB'}=f_\mathbb{X}^\mathbb{AB}=f_\mathbb{X}^\mathbb{A'B}=f_\mathbb{X}^\mathbb{A'B'}=f_\mathbb{X}$ independent of the arithmetics of Alice and Bob. In the first case
the CHSH random variable
\be
f_\mathbb{R}^\mathbb{AB}{}^{-1}(ab)+f_\mathbb{R}^\mathbb{AB'}{}^{-1}(ab')
+
f_\mathbb{R}^\mathbb{A'B}{}^{-1}(a'b)-f_\mathbb{R}^\mathbb{A'B'}{}^{-1}(a'b')
\ee
cannot be further simplified in general. 
In the second case the random variable is $f_\mathbb{X}^{-1}(\pm 2)$, but in general differs from $\pm 2$. The available structures are here much richer than in the standard approach. 
The slogan `Bell's theorem is a mathematical theorem' appears here in all its weaknesses and limitations. 

Finally, thinking in quantum cryptographic terms, how to formulate a test for eavesdropping? The intermediate arithmetics that define products of the results obtained by Alice and Bob involve additional arithmetics associated with the process of combining the data. This additional arithmetic may, in principle, involve an eavesdropper (or hacker) Eve.

\section{Remarks on non-classical correlations beyond quantum mechanics}

Examples where Bell-type inequalities are violated beyond quantum mechanics can be found in the literature. 
Which loopholes of the proof are there employed? 

In the model invented by Diederik Aerts \citep{Aerts1986}, after the first measurement of a random variable $A$ a hidden variable $\lambda$, an argument of $B(\lambda)$, is  found with nonzero probability in a state whose probability would be zero if  $A$ had not been measured. The measurement of $A$ nontrivially and `actively' changes the state of $B$. This is not the standard conditioning by `getting informed'. One should not confuse this type of conditioning with the one that leads to the Borel paradox \citep{MC92}. The model avoids the Borel paradox for the price of nonlocality in the sense of Bell. 
The example explains why nonclassical probabilities can occur in classical systems, but it does not contradict the conclusion of Bell. The other examples studied by the Brussels group, such as the two connected vessels of water \citep{vessels}, `Bertlmann wearing no socks' \citep{socks}, or Sven Aerts' mechanical model that simulates a non-local PR box \citep{Sven}, are also nonlocal in this sense. In all these cases a state $\lambda$ in $B(\lambda)$ is actively influenced by the choice of $A$ or $A'$ in the correlated system, and the mechanisms responsible for these influences are explicitly described. Information about the choice made by Alice has enough time to reach Bob, even though one cannot use this information to communicate. Action at-a-distance is present, but there is nothing `spooky' about it.

In another class of models the results $A$ and $B$ are created at the very moment the product $AB$ is computed. This is what happens in various cognitive-science examples where $B$ becomes a context for $A$ \citep{cogn0,cogn3}; see also the polemic \citep{D,H,A}. Contextuality of cognitive models is extensively studied, both theoretically and experimentally,  also in the works of \cite{Kh}. Exactly the same effect occurs in the rock-paper-scissors game, which is ubiquitously  present in many natural systems \citep{FS}. `Rock' in itself does not have a value $+1$ (win) or $-1$ (lose). However, the pairs (rock,paper) = $(-1,+1)$ and (rock,scissors)  = $(+1,-1)$ are uniquely defined. All these models are nonlocal in the sense of Bell, but there is nothing unphysical about their nonlocality. 

The controversial model by Christian \citep{JCh} is also explicitly based on the fact that there is a nontrivial relation between $A$, $B$, and $AB$.  Tensor product is there replaced by geometric product, a possibility considered in similar contexts also by other authors \citep{Doran,Doran2,AC2006,C2007,AC2008}). Operators such as Pauli matrices are interpreted in the standard geometric-algebra way, namely as a matrix representation of an orthonormal basis in $\mathbb{R}^3$ \citep{Hestenes}, so the whole calculation can be performed in purely geometric terms. On a philosophical side, there is some similarity between the way Christian treats binary variables with the one I have described above in the context of $1_\mathbb{X}$ and $\ominus_\mathbb{X}1_\mathbb{X}$. For Christian, `plus' and `minus' denote some kind of abstract handedness, defined at a quaternionic level. However, in spite of all his efforts, I believe the probability interpretation of multivector random variables is not yet sufficiently understood, a fact obscuring the physical meaning of the result. 

What I find most intriguing and worthy of further studies  is the link with the original model of \cite{Pearle}. Pearle's hidden variable is essentially an element of the rotation group SO(3), whereas Christian works with  SU(2). In both cases the hidden variables are classical but noncommutative (rotations in three dimensions do not commute). The random variables of Pearle are three-valued (0 and $\pm1$), where 0 is interpreted as no detection. So, Pearle's model formally looks like a spin-1 system. The model of Christian is essentially a spin-1/2, where the 0 eigenvalue is missing. Does it mean, as suggested by Christian, that undetected signals can be in principle eliminated if one considers hidden-variables that belong to the covering space of the Pearle model? In principle, it cannot be excluded because many structures occurring in quantum information theory have their space-time analogues (e.g. the two-qubit Bell basis is formally identical to the Minkowski tetrad \citep{Czachor2008}). This whole new research area is basically unexplored.

\section{Summary}

Bell's theorem cannot be proved if complementary measurements have to be represented by random variables that cannot be added or multiplied. Complementarity means here that there exists a fundamental logical reason why one cannot perform the two measurements simultaneously. In quantum mechanics one represents complementarity by noncommutativity. Alternatively, two quantum random variables are complementary if eigenvalues of their sum are not the sums of their eigenvalues. There is no proof that analogous properties cannot be found in non-quantum systems, a fact of fundamental importance for proofs of security in quantum cryptography. For example, random variables with values in non-Diophantine arithmetics are analogous to quantum observables belonging to different maximal sets of simultaneously measurable quantities.  Similarly to eigenvalues of non-commuting operators, it is not allowed to naively add or multiply elements from different non-Diophantine arithmetics, even if the arithmetics are defined in subsets of $\mathbb{R}$. Non-Diophantine arithmetics provide the first examples of non-quantum models where elements of reality in the sense of the EPR paradox are present, but the resulting complementary random variables cannot be added or multiplied, so that CHSH-type random variables cannot be automatically constructed. What I have shown in the paper is just a very preliminary study that should evolve into a larger research project.

\end{document}